\documentclass[aps,prl,superscriptaddress,floatfix,twocolumn,showpacs]{revtex4}

\usepackage{verbatim}

\usepackage{amsmath}
\usepackage{amsfonts}
\usepackage{amssymb}
\usepackage{graphicx}
\usepackage{bm}
\usepackage{color}
\usepackage{epsf}
\usepackage[hypertex]{hyperref}

\begin{document}
\title{
Boundary Multifractality at the Integer Quantum Hall Plateau
Transition:\\ Implications for  the Critical Theory}

\author{H. Obuse}
\altaffiliation[Present address: ]{Department of Physics,
Kyoto University, Sakyo-ku, Kyoto 606-8502, Japan}
\affiliation{Condensed Matter Theory Laboratory, RIKEN, Wako,
Saitama 351-0198, Japan}
\author{A. R. Subramaniam}
\affiliation{James Franck Institute and Department of Physics,
University of Chicago, Chicago, IL 60637, USA}
\author{A. Furusaki}
\affiliation{Condensed Matter Theory Laboratory, RIKEN, Wako,
Saitama 351-0198, Japan}
\author{I. A. Gruzberg}
\affiliation{James Franck Institute and Department of Physics,
University of Chicago, Chicago, IL 60637, USA}
\author{A. W. W. Ludwig}
\affiliation{Department of Physics, University of California,
Santa Barbara, CA 93106, USA}

\begin{abstract}

We study multifractal spectra of critical wave functions at the
integer quantum Hall plateau transition using the
Chalker-Coddington network model. Our numerical results provide
important new constraints which any critical theory for the
transition will have to satisfy. We find a non-parabolic
multifractal spectrum and determine the ratio of boundary to
bulk multifractal exponents. Our results rule out an exactly
parabolic spectrum that has been the centerpiece in a number of
proposals for critical field theories of the transition. In
addition, we demonstrate analytically exact parabolicity of
related boundary spectra in the two-dimensional chiral
orthogonal `Gade-Wegner' symmetry class.
\\[-0.5cm]
\end{abstract}


\date{August 11, 2008}
\maketitle

The physics of the  quantum Hall effect has been an exciting
area of research for more than two decades
\cite{vonKlitzing,QHeffects}. While much  progress has been
made in this area, the identification of an analytically
tractable theory describing the critical properties at the
transitions between the plateaus in the integer quantum Hall
(IQH) effect has been elusive ever since
\cite{PruiskenKhmelnitskii}. These quantum phase transitions
are famous examples of (Anderson) localization-delocalization
(LD) transitions driven by disorder. The diverging localization
length plays the role of a correlation length in non-random
continuous phase transitions, known to be described by
conformal field theories in two dimensions (2D). It is natural
to expect that effective (field) theories describing IQH
plateau transitions should generally also possess conformal
symmetry (cf.\ \cite{obuse2007}).

Many attempts have been made in the past to identify an
analytically tractable description of the IQH plateau
transition and, more recently, Wess-Zumino (WZ) field theories
defined on a certain supermanifold were  conjectured to provide
such a description \cite{zirnbauer,tsvelik,LeClair}. (Similar
theories have also appeared in the context of string
propagation in Anti-de Sitter space-time \cite{Strings}.) These
proposals focussed solely on bulk observables, i.e., on
physical quantities measured in a sample without any
boundaries. In this Letter, we provide important new
constraints that arise when one studies the scaling behavior of
wave functions near the boundaries of a sample. Any proposed
candidate theory for the plateau transitions will have to be
consistent with our numerical results for the boundary
multifractal spectrum.

At LD transitions, critical wave functions obey
scale-invariant, multifractal (MF) statistics, namely,
disorder-averaged moments of wave functions have a power-law
dependence on the linear dimension $L$ of the system
\cite{evers_review}:
\begin{equation}
\overline{|\psi(\bm{r})|^{2q}}\big/
\big(\overline{|\psi(\bm{r})|^2}\big)^q
= C^x_q(L) L^{-\Delta^x_q}.
\label{Delta_q_scaling}
\end{equation}
The MF exponents $\Delta^x_q$, which are related to
(`anomalous') scaling dimensions of certain operators in an
underlying field theory \cite{DuplantierLudwig1991}, can be
defined for points $\bm{r}$ in the bulk ($x = b$) of the sample,
$\Delta^b_q$, or near its boundary (`surface': $x=s$)
\cite{subramaniam,obuse2007}, $\Delta^s_q$. The prefactor
$C^x_q(L)$ in Eq.\ (\ref{Delta_q_scaling}) depends on $q$ and,
in general, on $L$ if we include the possibility of corrections
to scaling. Both sets of MF exponents satisfy the symmetry
relation \cite{mirlin06}
\begin{align}
\Delta^x_q = \Delta^x_{1-q}
\label{eq:symmetry}
\end{align}
(in some interval \cite{obuse2007} around $q=1/2$).

Equivalently, the MF wave functions can be characterized by the
so-called singularity spectra $f^x(\alpha^x)$ related to
$\Delta^x_q$ by a Legendre transform: $f^x(\alpha^x_q) =
(\alpha^x_q - 2)q - \Delta^x_q + d_x$, $\alpha^x_q - 2 = d
\Delta^x_q/dq$, and $d_b = 2$, $d_s = 1$. The exponent
$\alpha^x_0$ describes the scaling of typical wave functions:
$\overline {\ln |\psi(\bm{r})|^2} \sim - \alpha^x_0 \ln L$, as
can be seen by taking the $q$ derivative in Eq.\
(\ref{Delta_q_scaling}) at $q=0$.

Work emerging \cite{klesse01,evers2001} from Ref.\
\cite{zirnbauer} led to the conjecture that the proposed theory
would give rise to an exactly parabolic bulk MF spectrum for
the IQH transition
\begin{align}
\Delta^b_q = \gamma^b q(1-q),
\label{parabola}
\end{align}
reminiscent of analytically obtained MF spectra for Dirac
fermions in, e.g.,  random abelian  gauge potentials
\cite{ludwig94,mudry96}. In those models the parabolicity of
the MF spectrum can be understood through a reformulation of
the problem in terms of free fields.

Previous numerical studies \cite{evers2001} of wave function
statistics at the IQH transition appeared to exhibit a bulk MF
spectrum that was indeed well described (with an accuracy of
$\sim$1\%) by a parabolic fit (\ref{parabola}) with $\gamma^b =
0.262 \pm 0.003$, seemingly providing support for the
conjectures advanced in Ref.\ \cite{zirnbauer,tsvelik,LeClair}.
(In Ref.\ \cite{evers2001} the results are presented in terms
of $f^b(\alpha)$. For a parabolic MF spectrum (\ref{parabola}),
$f^b(\alpha^b)$ is also parabolic, with a maximum at
$\alpha^b_0 = \gamma^b + 2$.)

Besides its conjectured relevance \cite{zirnbauer} to the IQH
transition, the  above-mentioned WZ theory is known to describe
transport properties of a disordered electronic system in a
different universality class \cite{GadeWegner,GLL} (the chiral
unitary `Gade-Wegner' class AIII of
\cite{ZirnbauerJMP1996,altland}) which possesses an additional
discrete (chiral) symmetry \cite{ZirnbauerJMP1996}, not present
in microscopic models for the IQH transition. Well-known
microscopic realizations of field theories in class AIII are
random bipartite hopping models, and certain network models
\cite{GLL,GadeWegner,ChalkerBocquetConfirmsGLL}. The theory
possesses a line of fixed points, with continuously varying
critical properties parametrized by the critical longitudinal
DC conductivity. (It was argued in Ref.\ \cite{zirnbauer} that
for a particular value of this continuous parameter the WZ
theory would provide a description of the IQH transition.)

In this paper we obtain two kinds of results. First, we provide
results of extensive numerical work on the MF exponents at the
IQH transition both at a boundary ($\Delta_q^s$) and in the
bulk ($\Delta_q^b$). Based on these numerical results quadratic
behavior in $q$ is ruled out for both quantities. Deviations
from the parabolic form (\ref{parabola}) are found to be much
larger in the MF exponents $\Delta_q^s$ at a boundary. Here it
is important to note that in complete analogy to the bulk, the
above conjectures would {\it also} yield a quadratic dependence
on $q$ of the {\it boundary} MF exponents $\Delta^s_q$. We
further determine the ratio $\Delta_q^s$/$\Delta_q^b$ over a
range of $q$. Accounting for this ratio is an important
constraint on any proposed critical theory for the transition.

Secondly, we demonstrate analytically the exact parabolicity of
boundary spectra, not for the chiral unitary class AIII, but
for the related time-reversal invariant version, the chiral
orthogonal `Gade-Wegner' class BDI
\cite{GadeWegner,GLL,ChalkerBocquetConfirmsGLL,ZirnbauerJMP1996}.
We expect such parabolicity to also hold in the chiral unitary
symmetry class.

We begin with the numerical part. Here, we study the
multifractality of critical wave functions in a way similar to
Ref.\ \cite{evers2001}, with the goal of numerically
determining the rescaled anomalous exponents
\begin{align}
\gamma^x_q = \Delta^x_q/q(1-q),
\end{align}
both for $x = s$ (boundary) and $x = b$ (bulk).

For the case of boundary exponents we consider the critical
Chalker-Coddington network model (CCNM) \cite{chalker88} with
$4L^2$ links placed on a cylinder. The dynamics of wave
functions on links of the network is governed by a unitary
evolution operator $U$. For each disorder realization, we
numerically diagonalize $U$ and retain one critical wave
function whose eigenvalue is closest to 1. The largest system
size we studied was $L=180$, and the ensemble average was taken
over $3\times10^5$ samples for $L=50,60$, $5\times10^5$ samples
for $L=80$, and $2\times10^5$ samples for $L=120, 180$.

\begin{figure}
\centering
\includegraphics[width=0.45\textwidth]{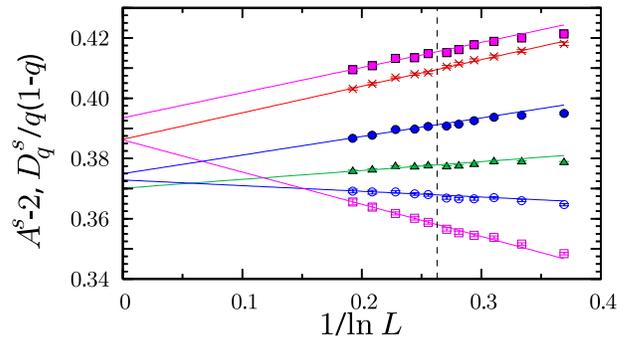}
\vspace*{-0.2cm}
\caption{
(Color online) System size dependence of $D^s_q(L)/q(1-q)$
defined in the main text for $q=-0.2$ ($\blacksquare$), $0.2$
($\bullet$), $0.5$ ($\blacktriangle$), $0.8$ ($\circ$), and
$1.2$ ($\square$). $\gamma^s_q$ is calculated by linear fitting
taking into account only for larger system sizes
($L\geqslant50$) indicated by the vertical dashed line between
$L=40$ and $50$. We also show $A^s - 2$ defined in the main
text ($\times$).} \label{fig:WFA} \vspace*{-0.4cm}
\end{figure}

We obtain the anomalous dimensions $\Delta^s_q$ from Eq.\
(\ref{Delta_q_scaling}). The boundary wave function
coarse-grained over each plaquette along the boundary, is
substituted into the left-hand side of Eq.\
(\ref{Delta_q_scaling}), where the overline denotes ensemble
and spatial averages along the boundary. Taking the logarithm,
we numerically obtain $D^s_q(L) \equiv (q \ln
\overline{|\psi(\bm{r})|^2} -\ln
\overline{|\psi(\bm{r})|^{2q}})/\ln L = \Delta_q^s - \ln
C^s_q(L)/\ln L$, and plot this quantity as a function of $1/\ln
L$ in Fig.\ \ref{fig:WFA}. We see that corrections to scaling
are significant for small systems ($L \lesssim 30$). Therefore,
we used our numerical data for $L \geqslant 50$ only to extract
$\gamma^s_q$ by linear fitting.

Independently, we numerically obtain $\alpha_q^s$ and $f^s$
from $ \overline{|\psi(\boldsymbol{r})|^{2q}
\ln|\psi(\boldsymbol{r})|^2} / \overline{
|\psi(\boldsymbol{r})|^{2q}} \sim - \alpha_q^s\ln L, $ and $
\ln\overline{|\psi(\boldsymbol{r})|^{2q}} \sim\left[
f^s(\alpha_q^s) - \alpha_q^s q - d_s \right]\ln L, $ using our
numerical data for $L\geqslant50$. For example, the exponent
$\alpha_0^s$ is obtained by linear fitting
(Fig.~\ref{fig:WFA}), $A^s \equiv -\overline{\ln
|\psi(\bm{r})|^2}/\ln L = \alpha_0^s + \mathrm{const.}/\ln L$,
which yields $\alpha^s_0 = 2.386 \pm 0.004$.

\begin{figure}
\centering
\includegraphics[width=0.45\textwidth]{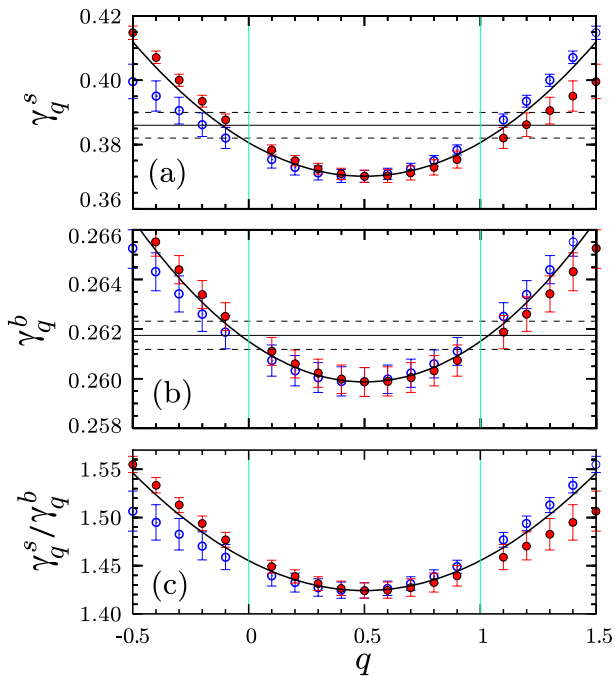}
\vspace*{-0.2cm}
\caption{
(Color online) (a) Rescaled boundary MF exponents $\gamma_q^s$
($\bullet$) and $\gamma_{1-q}^s$ ($\circ$). The curve is $0.370
+ 0.042 (q-1/2)^2,$ obtained by fitting the data for $\gamma^s_q$ in $0 < q <
1$ to a parabolic form. The horizontal solid line shows
$\alpha_0^s - 2 = 0.386 \pm 0.004$ with error bars indicated by
dashed lines, which is consistent with $\lim_{q\to0,1}
\gamma_q^s$. (b) Rescaled bulk MF exponents $\gamma_q^b$
($\bullet$) and $\gamma_{1-q}^b$ ($\circ$). The curve is
$0.2599 + 0.0065(q-1/2)^2$ obtained by fitting the data for $\gamma^b_q$ in $0
< q < 1$ to a parabolic form. The horizontal solid line shows
$\alpha_0^b - 2=0.2617 \pm 0.0006$ with error bars indicated by
dashed lines. (c) Ratios $\gamma_q^s/\gamma_q^b$ ($\bullet$)
and $\gamma_{1-q}^s/\gamma_{1-q}^b$ ($\circ$). As above, the curve is
obtained from the parabolic fits for $\gamma_q^{s,b}$, which
amounts to quartic approximations for $\Delta_q^{s,b}$. }
\label{fig:dimension} \vspace*{-0.4cm}
\end{figure}

We show in Fig.~\ref{fig:dimension}(a) the rescaled boundary
anomalous dimension $\gamma_q^s$ (red filled circles) obtained
from this analysis. We see clearly that $\gamma_q^s$ is not
constant, implying that the boundary MF spectrum $\Delta^s_q$
is {\it not parabolic}. The change in $\gamma_q^s$ over the
interval $0<q\leqslant1/2$ is about $4\sim5\%$ and is
significantly larger than the error bars. This provides the
strongest numerical evidence against the parabolicity of the MF
exponents.

Shown in the same figure by blue open circles is the mirror
image of $\gamma^s_q$ with respect to $q=1/2$,
$\gamma_{1-q}^s$. We see that the symmetry relation
(\ref{eq:symmetry}) is satisfied within error bars for $0
\lesssim q \lesssim 1$. The rescaled anomalous dimension
$\gamma_q^s$ approaches $\alpha_0^s - 2$ (the horizontal line)
at $q=0, 1$, indicating that the two independent calculations
of $\alpha_0^s$ and $\Delta_q^s$ are consistent.

We have also computed the bulk anomalous dimension $\Delta_q^b$
using the CCNM on a torus. In this case the overline in Eq.\
(\ref{Delta_q_scaling}) implies both the ensemble and the
spatial average over the whole torus. Wave functions are
coarse-grained on each plaquette. We have employed the same
fitting procedure as in the boundary case. The biggest system
size we examined for the bulk analysis is $L=270$. The number
of samples over which we took the average is $5\times10^5$ for
$L=50$, $3\times10^5$ for $L=60,80$, $2\times10^5$ for $L=120$,
$4\times10^4$ for $L=180$, and $2\times10^4$ for $L=270$.

Figure \ref{fig:dimension}(b) shows the exponents $\gamma_q^b$,
together with their mirror image. The symmetry relation
(\ref{eq:symmetry}) is again satisfied for $0\lesssim
q\lesssim1$ within error bars, which provides confirmation that
our results are reliable. We see clearly that $\gamma_q^b$ has
$q$ dependence, although it is weaker than that of
$\gamma_q^s$; compare the vertical scales of
Fig.~\ref{fig:dimension}(a) and (b).

The ratio $\gamma_q^s/\gamma_q^b$ is shown in
Fig.~\ref{fig:dimension}(c) and is seen to be clearly dependent
on $q$. Any candidate theory for the IQH transition needs to be
consistent with this ratio, and in particular its dependence on
$q$. (Note that for a free field this ratio would be equal to
2, and independent of $q$ \cite{Cardy1984,subramaniam}.)

\begin{figure}
\centering
\includegraphics[width=0.45\textwidth]{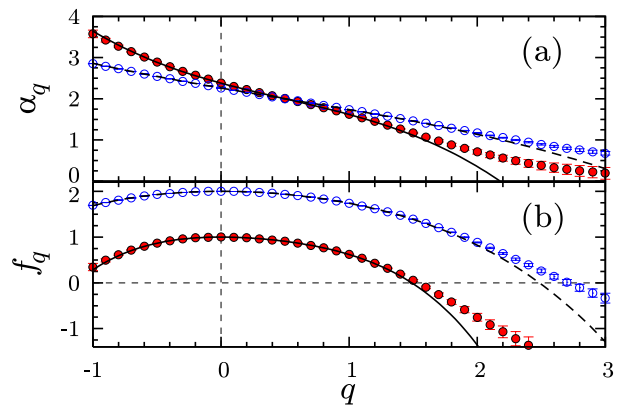}
\vspace*{-0.2cm}
\caption{
(Color online) (a) $\alpha_q^s$ ($\bullet$) and $\alpha_q^b$
($\circ$) as functions of $q$; (b) $f_q^s$ ($\bullet$) and
$f_q^b$ ($\circ$) as functions of $q$. The solid and dashed
curves on both panels are obtained from the parabolic
approximations to $\gamma_q^x$ (that is, quartic approximations
to $\Delta_q^x$). Notice that $\alpha^x_q$ significantly
deviate from straight lines which would follow from strictly
parabolic $\Delta^x_q$ (or constant $\gamma^x_q$).}
\label{fig:f-alpha}
\vspace*{-0.4cm}
\end{figure}

Figure \ref{fig:f-alpha}(a) shows $\alpha^x_q$ as a function of
$q$. The data significantly deviate from linear behavior that
would result if $\Delta^x_q$ were strictly parabolic (constant
$\gamma^x_q$). In Fig.\ \ref{fig:f-alpha}(b) we show the
corresponding singularity spectra $f^x(\alpha^x_q)$ as
functions of $q$. (Combining the data from the two panels would
result in $f^x(\alpha^x)$ as functions of $\alpha^x$.) For
$q\gtrsim1.5$ where $f^s(\alpha^s)<0$, the moments
$\overline{|\psi(\bm{r})|^{2q}}$ are dominated by rare events,
and thus accurate numerical calculation of MF exponents becomes
more difficult \cite{evers_review}.  This explains the
discrepancy between the (red) data points and the solid curves
for $q\gtrsim1.5$ in Fig.\ \ref{fig:f-alpha}. As $f_q^s>0$ at
$q\gtrsim-1$, we expect that our numerical results of
$f^s(\alpha^s)$ should be more reliable at $q\approx-1$ than at
$q\approx1.5$, as evidenced by the agreement between the red
dots and the solid curve in Fig.\ \ref{fig:f-alpha}. The curve
suggests termination of $f^s(\alpha^s)$ \cite{evers_review} to
occur at $q\approx2.2$.

We now proceed to present our second (analytical) result. We
first recall that the  theory conjectured in Ref.\
\cite{zirnbauer} to describe the IQH plateau  transition is a
WZ model with global psl(2$|$2) (super-) symmetry. This theory
possesses two coupling constants, one denoted by $f$ for the
kinetic term and another denoted by $k$ for the WZ term ($k=1$
in Ref.\ \cite{zirnbauer}), in standard notation
\cite{KnizhnikZamolodchikov,zirnbauer}. One can think of this
theory as a perturbation of the rather well understood
\cite{KnizhnikZamolodchikov,SchomerusQuellaWZW-On-PSU}
Kac-Moody (KM) point characterized by the condition $f^{-2} =
k$, perturbed by a term in the action of the form
\cite{KnizhnikZamolodchikov,QuellaSchomerusCreutzigDec2007}
$\delta S = (\lambda/k^2) \int d^2 z \, \phi_{ab}(z,{\bar z})
J^a(z) {\bar J}^b({\bar z})$, where $\lambda = f^{-2} - k$.
Here $\phi_{ab}$ is the KM primary field in the adjoint
representation of psl(2$|$2), $J^a$ and  ${\bar J}^b$ are the
left/right chiral components of the psl(2$|$2) Noether currents
\cite{KnizhnikZamolodchikov}, and $\lambda$ parametrizes the
line of fixed points, mentioned above.

The conjectured link \cite{zirnbauer}  between the WZ model and
the IQH transition can be formulated through  the notion of the
point contact conductance (PCC)
\cite{JanssenMetzlerZirnbauer1999}.  The PCC is a statistically
fluctuating quantity. (i) On one hand, the scaling dimension
$X_q$ of the $q$-th moment of the PCC  at the IQH transition
has been {\it proven} to be simply related to the exponent
$\Delta_q$ \cite{evers2001,klesse01}, $X_q = 2 \Delta_q$ ($|q|
< 1/2$). (ii) On the other hand, the scaling dimension $x_q$ of
the operator {\it in the WZ model} carrying the same
representation of the global psl(2$|$2) symmetry as the $q$-th
moment of the PCC in the CCNM (possessing the same psl(2$|$2)
symmetry), was {\it conjectured} \cite{zirnbauer} to be a
quadratic function of $q$. (iii) If one combines (i) and (ii),
and if one assumes $X_q = x_q$ (following the conjectured
description of the IQH transition by the WZ model), then the
wave function exponents $\Delta_q$ {\it at the IQH transition}
would be quadratic functions of $q$, as in Eq.\
(\ref{parabola}).

As already mentioned, this WZ theory is known to describe
transport properties of the chiral unitary class AIII
\cite{GadeWegner,GLL,ChalkerBocquetConfirmsGLL}, lacking
time-reversal symmetry. Below we demonstrate the correctness of
the conjecture  made in item (ii) above, at a {\it  boundary},
and for the {\it time-reversal invariant version} of the AIII
model,  the chiral orthogonal class BDI
\cite{GadeWegner,ZirnbauerJMP1996}. Just as its cousin with
broken time-reversal symmetry, the chiral orthogonal theory
also  possesses a line of fixed points. Transport properties
along this line can be described \cite{GLL} by  the
perturbation of the KM point of the psl(2$|$2)-invariant WZ
theory described above, when the field $\phi_{ab}$ is replaced
by the Kronecker delta, $\phi_{ab}$ $\to \delta_{ab}$. Denote
the corresponding coupling constant by $\lambda_t$. Consider
the theory in the upper half plane where the system simply ends
at the boundary, and an operator of scaling dimension (a
`conformal weight') $x^s_{\rho}(\lambda_t)$ on the boundary. At
the KM point, where the perturbation vanishes, $\lambda_t=0$,
such an operator is described by a representation $\rho$ of the
global psl(2$|$2) symmetry. It is known
\cite{KnizhnikZamolodchikov} that $x^s_{\rho}(\lambda_t=0)=$
$C^{(2)}_{\rho}/k$, where $C^{(2)}_{\rho}$ is the quadratic
Casimir invariant in the representation $\rho$. It turns out to
be straightforward
\cite{PerturbationQuellaSchomerusCreutzigDec2007} to compute
the change of the scaling dimension, order by order in the bulk
coupling constant $\lambda_t$, yielding a geometric series. The
result is simply $x^s_{\rho}(\lambda_t)=$
$C^{(2)}_{\rho}/(k+\lambda_t)$. Note that for the (continuous
series) representation $\rho$ of psl(2$|$2) in which the $q$-th
moment of the  PCC at the IQH transition transforms, one has
$C^{(2)}_{\rho} = q (1-q)$. This proves our claim that the
spectrum of scaling dimensions $x^s_\rho \to x^s_q$ of
corresponding boundary operators in symmetry class BDI is a
strictly quadratic function of $q$.

In summary, our numerical results clearly demonstrate that
both, the boundary and the bulk MF spectra, $\Delta^s_q$ and
$\Delta^b_q$, significantly deviate from parabolicity, and that
their $q$-dependent ratio is significantly different from 2.
(These conclusions were recently also reached, independently,
by Evers, Mildenberger, and Mirlin \cite{paper1}.) These
results for the  bulk as well as the boundary MF spectra impose
important constraints on any analytical theory for the IQH
plateau transition. Furthermore, we have demonstrated
analytically exact parabolicity of related boundary spectra in
the 2D chiral orthogonal `Gade-Wegner' symmetry class BDI.

We thank F. Evers, A. Mildenberger, and A. D. Mirlin for
helpful discussions and for sharing their data prior to
publication. We are grateful to T. Ohtsuki for his helpful
suggestions on numerical algorithm. This work was partly
supported by the Next Generation Super Computing Project,
Nanoscience Program and a Grant-in-Aid for Scientific Research
(No.~16GS0219) from MEXT, Japan (HO and AF), NSF Career award DMR-0448820, NSF
MRSEC DMR-0213745 (IAG), and DMR- 0706140 (AWWL). Numerical
calculations were performed on the RIKEN Super Combined Cluster
System.

\end{document}